# More Evidence for a Distribution of Tunnel Splittings in Mn$_{12}$-acetate

K. M. Mertes, Yoko Suzuki and M. P. Sarachik
*Physics Department, City College of the City University of New York, New York, NY 10031*

Y. Myasoedov, H. Shtrikman, and E. Zeldov
*Dept. of Condensed Mater Physics, The Weizmann Institute of Science, Rehovot 76100, Israel*

E. M. Rumberger and D. N. Hendrickson
*Dept. of Chemistry and Biochemistry, University of California at San Diego, La Jolla, CA 92093*

G. Christou
*Department of Chemistry, University of Florida, Gainesville, FL 32611*
(November 10, 2018)

In magnetic fields applied parallel to the anisotropy axis, the magnetization of Mn$_{12}$ has been measured in response to a field that is swept back and forth across the resonances corresponding to steps $N = 4, 5, ...9$. The fraction of molecules remaining in the metastable well after each sweep through the resonance is inconsistent with expectations for an ensemble of identical molecules. The data are consistent instead with the presence of a broad distribution of tunnel splittings. A very good fit is obtained for a Gaussian distribution of the second-order anisotropy tunneling parameter $X_E = -\ln(|E|/2D)$. We show that dipolar shuffling is a negligible effect which cannot explain our data.

Much attention has been focussed on the behavior of high-spin molecular nanomagnets because of their intrinsic interest as well as for their potential applications in high density storage of information and possible use as qubits for quantum computation. In the prototypical single molecule magnet Mn$_{12}$-acetate, ([Mn$_{12}$O$_{12}$(CH$_3$COO)$_{16}$(H$_2$O)$_4$]·2CH$_3$COOH·4H$_2$O), recent studies [1] of the magnetic relaxation at temperatures below 0.5 K for different sweep rates of a magnetic field applied along the longitudinal (easy) axis have yielded evidence that the symmetry breaking field that drives the tunneling in this material is locally varying, broadly distributed second-order anisotropy which is forbidden by the tetragonal symmetry of the Mn$_{12}$-acetate crystal, but which is present in real crystals due perhaps to long-range crystal dislocations or local chemical environment. The distributed second order transverse anisotropy gives rise to a distribution of tunnel splittings. In this paper, we present additional experimental evidence that the tunnel splittings are broadly distributed.

Mn$_{12}$-acetate is composed of weakly-interacting magnetic clusters, each consisting of twelve Mn atoms coupled by superexchange through oxygen bridges to give a net spin $S = 10$ at low temperatures. The magnetic molecules are regularly arranged on a tetragonal body-centered lattice with strong uniaxial anisotropy of the order of 60 K. Below the blocking temperature, $T_B \approx 3$ K, steep steps are observed [2] in the $M$ versus $H$ curves due to enhanced relaxation of the magnetization whenever levels on opposite sides of the anisotropy barrier coincide in energy. The steps are separated by plateaux where magnetic relaxation requires thermal activation and is strongly suppressed.

The Hamiltonian for Mn$_{12}$-acetate is given by:

$$\mathcal{H} = -DS_z^2 - AS_z^4 - g_z\mu_B H_z S_z + V_T, \quad (1)$$

where the anisotropy $D = 0.548(3)K$, the fourth-order longitudinal anisotropy $A = 1.173(4) \times 10^{-3}$ K, $g_z$ is estimated to be 1.94(1) [3], and $V_T$ contains all symmetry-breaking terms responsible for the observed tunneling. For temperatures below approximately 0.5 K, it has been shown that essentially all the tunneling proceeds from the ground state level $m' = -10$ of the metastable well [4]. Ground state tunneling, indicated by the arrow in the doublewell structure in Fig. 1, occurs at magnetic fields:

$$H_z = N\frac{D}{g_z\mu_B}\left[1 + \frac{A}{D}\left(N^2 - 20N + 200\right)\right], \quad (2)$$

where $N = -(m + m')$ is the step number. The data reported in this paper were obtained at 0.25 K, where only ground state tunneling is observed.

The magnetization of small single crystals of Mn$_{12}$-acetate was determined by methods described elsewhere [5]. The relaxation of the magnetization was studied for each observable ground state resonance as follows: starting from zero magnetization, the external field was swept from zero up to and beyond a given resonance $N$. When the field reached the middle of the plateau just above step $N$, it was reversed and swept back across this resonance. When the field reached the middle of the previous plateau, the field was swept back up and the process was repeated many times. The same procedure was repeated for all observed resonances, $N = 4, 5, ...9$, as shown in the normalized magnetization curves of Fig. 1. The turning points were chosen so that the amplitude of the field oscillations were the same for each $N$ and the turning points of the field occurred near the center of a plateau. The field was swept at a rate $dH_z/dt = 5.83$mT/s.

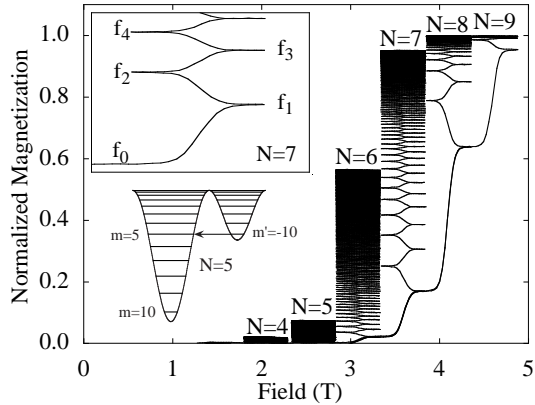

FIG. 1. Normalized magnetization versus magnetic field for a field that is swept back and forth across each resonance. The field was swept at a rate of $dH_z/dt = 5.83$mT/s. The upper inset shows the labeling described in the text for the first few oscillations for $N = 7$. The arrow in the lower inset designates tunneling from the ground state of the metastable well for step $N = 5$.

As the field is swept back and forth across the field resonance, the magnetization relaxes in steps. The manner in which it relaxes is expected to be different for a set of identical molecules than for the case of a distribution of tunnel splittings. We will show below that an ensemble of identical molecules cannot describe the data presented here. A method of fitting different distributions to the magnetization data will be developed, and an excellent fit is obtained for a lognormal Gaussian distribution. It will also be argued that dipolar shuffling [6] would produce an asymmetry in the steps which is not observed.

The fraction of molecules remaining after each step can be determined from the magnetization curves. At the turning points labeled by $f_0, f_1, f_2,...$, for $N = 7$, shown in the upper inset of Fig. 1, the fraction of molecules remaining in the metastable well is related to the normalized magnetization by,

$$f_j = (1 - M_j)/2, \qquad (3)$$

where $M_j$ is the normalized magnetization at the turning points of the field. For each $N$, Fig. 2 shows the fraction of molecules remaining in the metastable well after each field pass plotted as a function of how many times, $j$, the field was swept past the resonance. The issue is whether or not the fraction remaining after each field sweep is consistent with the behavior expected for a sample comprised of an ensemble of identical molecules.

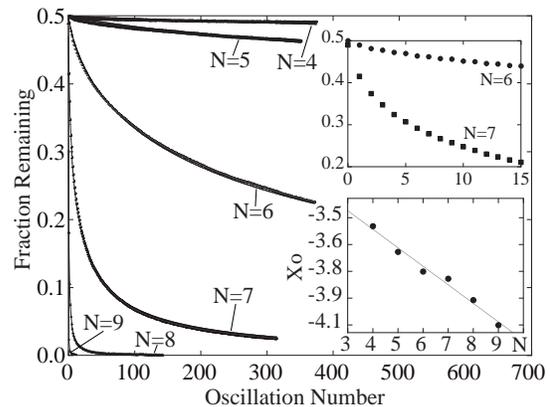

FIG. 2. Fraction of molecules remaining in the metastable well as a function of the oscillation number, $j$. With a Gaussian distribution of $X_E = -\ln(|E|/2D)$, the fits to Eq. 5 for each $N$ shown by the solid lines are nearly indistinguishable from the measured data. The upper inset shows the first few data points for $N = 6$ and 7 on a linear scale to illustrate that dipolar shuffling has minimal effects (see text). The lower inset shows the dependence of the fitting parameter, $X_0$, on $N$.

For a set of identical molecules, the probability, $P_N$, of remaining in the metastable well for resonance, $N$, is the same for each molecule. If the fraction of molecules in the metastable well before sweeping past the resonance is $f_{N,0}$, then the fraction of molecules remaining in the metastable well after sweeping the field through a resonance will be $f_{N,1} = f_{N,0}P_N$. When the field is swept back down across the resonance a second time, another set of molecules will tunnel. The fraction remaining after the second pass can be determined from the fraction remaining after the first pass, $f_{N,2} = f_{N,1}P_N = f_{N,0}(P_N)^2$. In general, the fraction of molecules remaining in the metastable well after $j$ passes of the field is:

$$f_{N,j} = f_{N,0}(P_N)^j \qquad (4)$$

This form implies that for a set of identical molecules, $f_{N,j}$ would depend exponentially on $j$, and a semi-log plot of the remaining fraction, $f_{N,j}$, versus the number of oscillations, $j$, should yield straight lines for each $N$. As can be seen in Fig. 3, which shows the data of Fig. 2 on a semi-logarithmic scale, this is not the case. The simple assumption that all the molecules are identical is clearly inconsistent with our data.

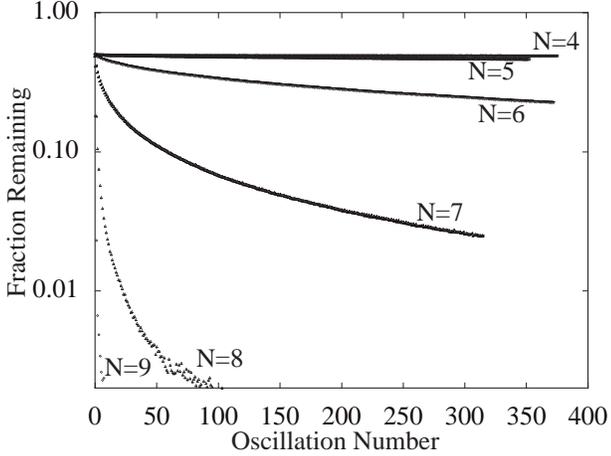

FIG. 3. The fraction of molecules remaining in the metastable well should have an exponential dependence on the oscillation number, $j$, for an ensemble of identical molecules (see text). Since the data is not linear on a semilogarithmic plot, our results are inconsistent with expectations for identical molecules.

We now consider the case of a distribution of tunnel splittings. Chudnovsky and Garanin [7] calculated the effect of such a distribution due to transverse anisotropies caused by crystal dislocations which produce deformations that require inclusion of a magnetoelastic term in the Hamiltonian. The dislocations give rise to a transverse anisotropy, $E(S_x^2 - S_y^2)$ that varies with the distance from the dislocation. Thus, the second order transverse anisotropy prefactor, $E$, is distributed. Alternatively, Cornia et al. [8] have pointed out that there are six different isomers of $Mn_{12}$-acetate, four of which are only twofold symmetric. This lowers the symmetry group for the Hamiltonian and also permits inclusion of locally varying second-order transverse anisotropy terms. In either case, a distribution of second-order anisotropy gives rise to a distribution of tunnel splittings. The following analysis is valid regardless of the origin of such a distribution of the second order transverse anisotropy prefactor, $E$.

The probability of remaining in the metastable well is given by the incoherent [9] Landau-Zener-Stueckelberg formula [10], $P_N = \exp(-\pi \Delta_N^2 / 2\hbar v_N)$, where $\Delta_N$ is the tunnel splitting and $v_N = (2S - N)g\mu_B dH_z/dt$ is the energy sweep rate. The tunnel splitting can be calculated through high order degenerate perturbation theory. For the case of second order transverse anisotropy (neglecting the fourth order longitudinal anisotropy), the tunnel splitting has the form, $\Delta_N = \eta_N g_N \left(\frac{|E|}{2D}\right)^{\xi_N}$, where $g_N = \frac{2D}{[(2S-N-2)!!]^2}\sqrt{\frac{(2S-N)!(2S)!}{(N)!}}$ and we set $D = 0.65$ K (instead of $D = 0.548$ K) to compensate for the neglect of the fourth order term. For even $N$, $\xi_N = S - N/2$ and $\eta_N = 1$. For odd $N$, $\xi_N = S - (N-1)/2$ and $\eta_N = N/2$. [7,5]

For an arbitrary initial distribution, $f_0(E)$, the fraction of molecules remaining in the metastable well with a transverse anisotropy prefactor between $E$ and $E + dE$ is $f_1(E) dE = P_1(E) f_0(E) dE$, where $P_1(E)$ is the probability of remaining in the metastable well for the $N = 1$ resonance. Integrating over all possible transverse anisotropy prefactors, one obtains the fraction of molecules remaining in the metastable well after having swept the field entirely through the resonance, $\tilde{f}_1 = \int_{-\infty}^{+\infty} P_1(E) f_0(E) dE$. The total fraction of molecules remaining in the metastable well after having swept the field back and forth across the $N = 1$ resonance $j$ times is, $\tilde{f}_{1,j} = \int_{-\infty}^{+\infty} (P_1(E))^j f_0(E) dE$. Note that for $N > 1$, the field is swept once through each of the lower resonances. Therefore, the fraction of molecules remaining in the metastable well, after sweeping the field back and forth $j$ times across resonance $N$ is given by,

$$\tilde{f}_{N,j} = \int_{-\infty}^{+\infty} \prod_{k=0}^{N-1} P_k(E) \left(P_N(E)\right)^j f_0(E) dE, \qquad (5)$$

where $\prod_{k=0}^{N-1} P_k(X_E)$ determines how much of the distribution remains after having swept through the previous $N - 1$ resonances and $P_N(E)$ is the probability of remaining in the metastable well for the $N^{th}$ resonance. Eq. 5 can be used to model the data shown in Fig. 2. Different distributions, $f_0(E)$, can be chosen and all the curves in Fig. 2 should be fit simultaneously for the same choice of parameters defining the shape of $f_0(E)$.

An excellent fit to the data is obtained for a log-normal distribution of the form [1] $f(X_E) = e^{-(X_E - X_0)^2/4\sigma^2}$, where $X_E = -\ln(|E|/2D)$. This form captures the behavior remarkably well for all observed resonances, and is indistinguishable from the data shown in Figs. 2 and 3. However, rather than being constant, the parameter $X_0$ required for a best fit was found to vary weakly with step number $N$. The values of $X_0$ chosen for each $N$ in the fit are shown in the inset to Fig. 2. The width of the distribution, $\sigma = 0.09814$, was constrained to be constant for the fit. The variation of $X_0$ with $N$ could be due perhaps to the existence of other symmetry-breaking terms in the Hamiltonian. In our previous studies [1], deviations from full scaling were attributed to a small admixture of tunneling due to transverse magnetic fields.

"Dipolar shuffling" has been recently proposed as an explanation for an unexpected dependence of tunnel splittings on sweep rate [6]. This effect refers to the fact that when a molecule tunnels, the internal dipole fields are locally altered so that some molecules that were previously below are now above the resonance field and vice-versa. For the long thin sample studied here ($a \approx 40\mu$m, $b \approx 40\mu$m, $c \approx 150\mu$m), this is expected to introduce an asymmetry between up- and down-sweeps: during an up-sweep many of the dipole-shuffled molecules will skip the resonant field resulting in a lower tunneling rate while during a down-sweep molecules will have additional opportunities to tunnel giving an enhanced tunneling rate. This would yield a tilted staircase behavior in the fraction of molecules with successive values of $j$. The upper inset of Fig. 2 shows that the fraction remaining varies smoothly with oscillation number and does not exhibit a tilted staircase behavior. We conclude that for the samples studied here, there was no observable effect associated with dipolar shuffling, which cannot therefore be responsible for the nonexponential behavior shown in Fig 3.

In summary, the dependence of the measured tunneling probability on the amount of magnetization remaining in the metastable well was studied by sweeping the field back and forth across each field resonance $N$. The fraction of molecules remaining in the metastable well after each sweep through the resonance was shown to be inconsistent with expectations for an ensemble of identical molecules. The data were analyzed on the assumption that there is a distribution of tunnel splittings associated with locally varying second-order transverse anisotropy arising from dislocations or a local symmetry-breaking term due to detailed chemical environment. A very good fit is obtained for a Gaussian distribution in $X_E = -\ln(|E|/2D)$. A variation of $X_0$ on $N$ is found which may be due to the existence of other symmetry-breaking terms in the Hamiltonian. In principle, the analysis presented here offers a method for determining the distribution of tunnel splittings.

Work at City College was supported by NSF grant DMR-0116808 and at the University of California, San Diego by NSF grants CHE-0095031 and DMR-0103290. Support for G. C. was provided by NSF grants DMR-0103290 and CHE-0123603. E. Z. and H. S. acknowledge the support of the Israel Science Foundation Center of Excellence grant 8003/02.